\documentclass[aps,prx,twocolumn,superscriptaddress]{revtex4}
\usepackage{amssymb}
\usepackage{amsfonts}
\usepackage{rotating}
 \usepackage{amsmath}
\usepackage{graphics}
\usepackage{palatino}
\usepackage{epsfig}
\usepackage[pdftex,usenames]{color}

\newcommand{\be}{\begin{equation}}
\newcommand{\ee}{\end{equation}}

\newcommand{\bea}{\begin{eqnarray}}
\newcommand{\eea}{\end{eqnarray}}

\begin{document}
% \draft
%%\large

%\title{And the winner is...Forecasting American Idol outcomes with Twitter}
%\title{And the winner is... beating the news on American Idol by using Twitter data analytics}
\title{Beating the news using Social Media: the case study of American Idol\footnote{This is an updated version of the paper where data gathered during the show and voting time of May 22 have been processed. The only changes to the manuscript are in Section 5 where we discuss the real time predictions concerning the season $11$ finale.}}

\author{Fabio Ciulla}
\affiliation{Department of Physics, College of Computer and Information Sciences, Department of Health Sciences, Northeastern University, Boston MA 02115 USA}

\author{Delia Mocanu}
\affiliation{Department of Physics, College of Computer and Information Sciences, Department of Health Sciences, Northeastern University, Boston MA 02115 USA}

\author{Andrea Baronchelli}
\affiliation{Department of Physics, College of Computer and Information Sciences, Department of Health Sciences, Northeastern University, Boston MA 02115 USA}

\author{Bruno Gon\c calves}
\affiliation{Department of Physics, College of Computer and Information Sciences, Department of Health Sciences, Northeastern University, Boston MA 02115 USA}

\author{Nicola Perra}
\affiliation{Department of Physics, College of Computer and Information Sciences, Department of Health Sciences, Northeastern University, Boston MA 02115 USA}

\author{Alessandro Vespignani}
\affiliation{Department of Physics, College of Computer and Information Sciences, Department of Health Sciences, Northeastern University, Boston MA 02115 USA}
\affiliation{Institute for Scientific Interchange Foundation, Turin 10133, Italy}
\affiliation{Institute for Quantitative Social Sciences, Harvard University, Cambridge, MA, 02138} 

\date{\today} \widetext

\begin{abstract} 
We present a contribution to the debate on the predictability of social events using big data analytics. We focus on 
the elimination of contestants in the American Idol TV shows as an example of a well defined electoral phenomenon that each week draws millions of votes in the USA.
We provide evidence that Twitter activity during the time span defined by the TV show airing and the voting period following it, correlates with the contestants ranking and allows the anticipation of the voting outcome. Twitter data from the show and the voting period of the season finale have been analyzed to attempt the winner prediction at 10.00 am of May the 23rd ahead of the airing of the official result. Furthermore, the fraction of Tweets that contain geolocation information allows us to map the fanbase of each contestant, both within the US and abroad, showing that strong regional polarizations occur. Although American Idol voting is just a minimal and simplified version of complex societal phenomena such as political elections, this work  shows that  the volume of information available in online systems permits the real time gathering of quantitative indicators anticipating the future unfolding of opinion formation events.
\end{abstract}

\maketitle

The recent global surge in the use of technologies such as Social Media, smart phones and GPS devices  has changed the way in which we live our lives in a fundamental way. Our use of such technologies is also having a much less visible, but not less significant, consequence: the collection on a massive scale of extremely detailed data on social behavior is providing a unique and unprecedented opportunity to observe and study social phenomena in a completely unobtrusive way. The public availability of such data, although limited, has already ignited a flurry of research into the development of indicators that can act as distributed proxies for what is occurring around the world in real time. In particular,  search engine queries or posts on microblogging systems such as Twitter have been used to forecast epidemics spreading \cite{culotta2010towards}, stock market behavior \cite{Bollen20111} and election outcomes\cite{tumasjan2010predicting,livne2011party,skoric2012tweets,sang2012predicting} with varying degrees of success. 
However, as many authors have pointed out, there are several challenges one must face when dealing with data of this nature:  intrinsic biases, uneven sampling across location of interest etc.~\cite{ratkiewicz2011detecting,metaxas2011not,mislove2011understanding,gayo2012wanted}.

In this paper we intend to assess the usefulness of open source data by analyzing in depth the microblogging activity surrounding the voting behavior on the contestants in American Idol, one of the most viewed American TV Shows. In this program, the audience is asked to choose which contestant goes forward in the competition by voting for their favorites. The well delineated time frame (a period of just a few hours) and frequency (every week) over an extended period (an entire TV Season) provides a close to ideal test ground for the study of electoral outcomes as many of the assumptions implicitly used in the analysis of social phenomena are more easily arguable, if not trivially true, in the case of the American idol competition. In particular, we assume that:

\begin{itemize}
\item The demographics of users tweeting about American Idol are representative of the voting pool.
\item The self-selection bias, according to which the people discussing about politics on Twitter are likely to be activists scarcely representative of the average voter, seems to become almost a positive discrimination factor in the case of a TV show where the voters are by definition self-selected. 
\item Voting fans are the most motivated subset of the audience (the population we are trying to probe) that are willing to make an extra effort for no personal reward, and, crucially, they are allowed to vote multiple times. 
\item Users are not malicious, and engage only in conversations he or she has a particular interest in
\item The influence incumbency, which strongly affects the outcome of political elections, is not a factor determining the outcome of American Idol.
\end{itemize}

For the above reasons we can consider TV show competitions as a case study for the use of open source indicators to achieve predictive power, or simply beating the news, about social phenomena. It is thus not surprising that other attempts to use open source indicators in this context have been proposed in the past. Here however we benefit from the  the constant growth of Twitter that makes it easier to collect significative statistical sample of the population. Furthermore, TV shows are now leveraging on Twitter and other social platform which are becoming in all respects a mainstream part of the show. This amplifies the importance of the indicators one can possibly extract from these media in monitoring the competition.

\section{Rules and voting system}

The first episode of the $11^{th}$ season of American Idol was aired on January $18$, $2012$ with a total of $42$ contestants. After an initial series of eliminations made by the judges, a final set of $13$ participants was selected. All further eliminations were decided by the audience through a simple voting system. During this final phase of the competition, two episodes are aired each week: On Wednesday the participants perform on stage and the public is invited to vote for two hours after the show ends. Voting can take one of three forms: toll-free phone calls, texting and online voting. The rules of the competition only allow for votes casted by the residents of the U.S., Puerto Rico and U.S. Virgin Islands. There is no limit to the number of messages or calls each person can make, while the online votes are limited to $50$ per computer as identified by its unique IP address. Every week, hundreds of millions of votes are counted and the contestant that gathers the least number is eliminated.
The show airs at $8.00$ PM local time on each coast. As a result of the time zone difference of three hours between the East and West coasts, the total voting window between the first and last possible vote is  $10.00$PM-$3.00$AM EST. During the season's final performance episode the voting window is extended to four hours after the show airs, resulting in an extended voting window between $10.00$PM-$5.00$AM EST.

\section{Data}

Our fundamental assumption is that the attention received by each contestant in Twitter is a proxy of the general preference of the audience. To validate this assumption, we collected tweets containing a list of $51$ \#tags, \@usernames and strings related to the show. 
The main dataset was obtained by extracting matching tweets from the raw Twitter feed used by Truthy~\cite{ratkiewicz11-2} for the entire duration of the current season of American Idol. The feed is a sample of about $10\%$ of the entire number of tweets that provides a, statistically significant, real time view of the topics discussed within the Twitter ecosystem. This allowed us to make a post-event analysis of the last $9$ eliminations. This dataset was further complemented by the results of automatically querying the Twitter search API every $10$ minutes for tweets containing one or more of the keywords we identified as related to American Idol. The search API data cover the period since May $16$, giving us a more detailed view of the last elimination before the season's finale. 

\section{A cartography of the fanbase}

\begin{figure}
\includegraphics[width=0.9\columnwidth,angle=0]{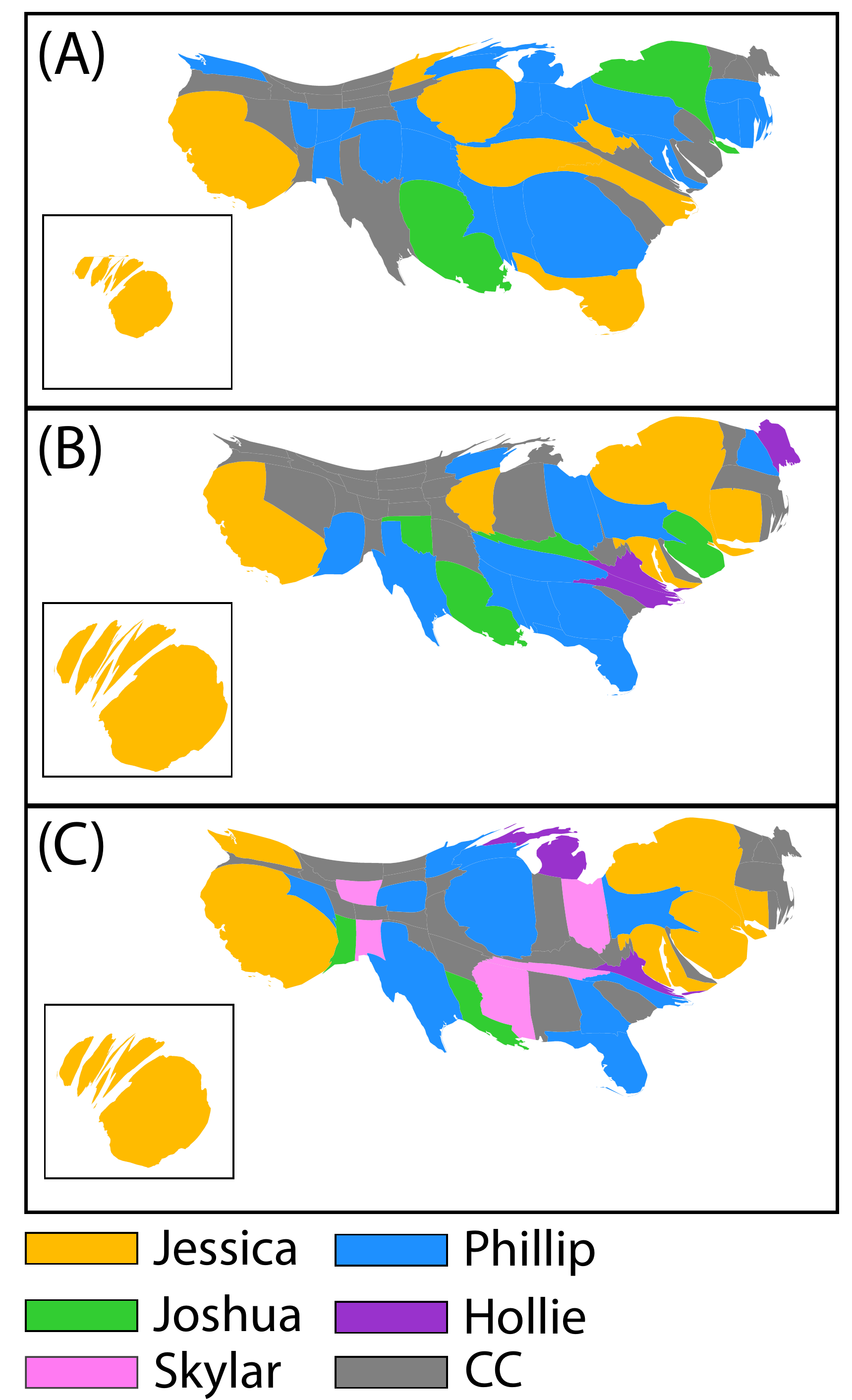}
\caption{Geographical polarization of the tweets for the Top $3$ (A), Top $4$ (B) and Top $5$ (C) episodes. The area of each State is proportional to the number of geolocalized tweets generated there, while the color represents the contestant with the majority of the vote. The grey represent states we could not assign to a single contestant within the statistical errors (CC). }
\label{cartogram}
\end{figure}

Tweets in our dataset often contain georeferenced location information that allows us to analyze the spatial patterns in voting behavior. Figure \ref{cartogram} shows a strong geographical polarization in the U.S. towards different candidates. In the weeks preceding the Top $3$ show [panels (B) and (C)], for example, Phillip Phillips gathers most of the attention in the Midwest and South, while Jessica Sanchez appears to be popular particularly on the West Coast as well as in the large metropolitan areas across all of the country, and Joshua Ledet is strong in Louisiana. The Top $3$ week analysis [panel (A)] shows a disturbance from the previous geographical distribution, perhaps due to the performance of the candidates. As expected, the audience reacts to the events occurring on Wednesday night. On the other hand, and perhaps not surprisingly, the attention basins of each of the three participants always include their origin city (Phillips was born and raised in Georgia, Sanchez is from Chula Vista, California, and Ledet from the Lake Charles metropolitan area in Louisiana)\footnote{http://www.americanidol.com/contestants/season\_ 11}. 
 
\begin{table}[t]
    \begin{tabular}{|c|c|c|c|}
    \hline
    { \bf Contestant} & { \bf U.S.A. } & { \bf World} & { \bf Philippines} \\ \hline %& 
%    $l_e$ & $\langle l \rangle$ & $ \langle \Delta t_l \rangle$ &
%    $\langle l_{proj} \rangle$ \\ \hline %old <s>
      Jessica &  $45 \pm  4 $ & $  64.2 \pm 2.2 $ & $92.8 \pm 1.9 $ \\ \hline
      Joshua & $ 15 \pm 3  $ & $9.8 \pm 1.3 $ & $1.4 \pm 0.9 $ \\ \hline
      Phillip & $40 \pm  4 $ &  $ 26 \pm 2.0  $ & $ 5.8 \pm 1.7 $\\ \hline
    \end{tabular}
  \caption{Popularity basins. Data concerns the entire American Idol season up to the morning of May $17$ (before the two finalists were announced), and refers to the percentage ($\%$) of popularity within U.S., the whole World and the Philippines. The geo-localized database for the three candidates contains $3251$ data points. Errors represent the normal confidence interval with a confidence level of $99\%$. } 
  \label{tab:geoloc}
\end{table}

\begin{figure}
 \includegraphics[width=0.9\columnwidth,angle=0]{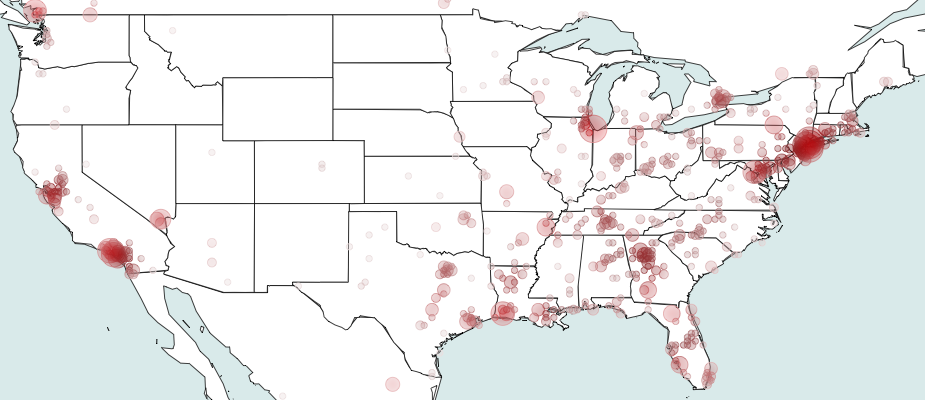} \\ \vspace{0.2cm}
\includegraphics[width=0.9\columnwidth,angle=0]{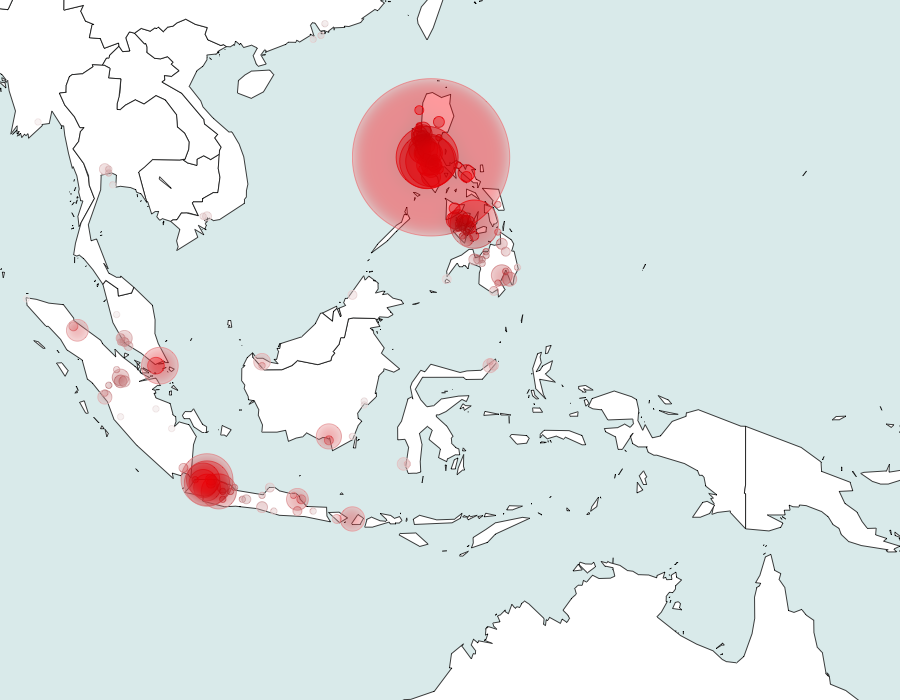} 
\caption{Local and global attention towards American Idol. Top: U.S. data show that the highest Twitter activity is concentrated in the large metropolitan areas, as expected. Bottom: The Philippines are distinctly more active  than any other foreign Country. It is worth noting also that a remarkable signal is produced in Indonesia, too, which is very active country with respect to Twitter activity in general.}
\label{fig:world}
\end{figure}

The geolocalized data also allows for a unique view of the attention devoted to American Idol in the rest of the world.
Although one might naively expect interest to be limited to the US, Figure \ref{fig:world} shows that the show is also popular in several foreign countries and particularly in the Philippines. This can be understood by noting that one of the contestants is of Filipino origin. Jessica Sanchez's mother is originally Filipino, having been born in the Bataan province \footnote{http://newsinfo.inquirer.net/183767/jessica-sanchez-makes-it-to-idols-top-5-rocker-elise-testone-out}. Participation in American Idol has made Sanchez so popular in her mothers native country that on May $16$ the Philippine President Benigno Aquino III congratulated the singer for her performance and stated, ``Hopefully she really reaches the top.'' \footnote{http://www.washingtonpost.com/entertainment/tv/\\philippine-president-aquino-roots-for-jessica-sanchez-to-win-american-idol-next-week/2012/05/16/gIQAwsWzSU\_story.html}. Table \ref{tab:geoloc} quantifies this intuition. Jessica Sanchez related Tweets are $45\%$ of the total if only U.S. is considered, while it rises to $64\%$ if the whole World is considered.  Officially,  Sanchez's popularity abroad should not have any impact on voting, since, as mentioned above, only the U.S. based audience is allowed to take part into the election procedure. However, it is interesting to note that the Filipino-restricted Twitter activity concerning Jessica is strongly peaked in the two voting sessions of American Idol for the East and West timezones, and that numerous websites explicitly address the issue of "voting tunnels": ``How to Vote for Jessica Sanchez from the Philippines and Other Non-US Countries'' \footnote{http://www.starmometer.com/2012/04/19/how-to-vote-for-jessica-sanchez-from-the-philippines-and-other-non-us-countries/, http://www.gmanetwork.com/news/story/258482/\\pinoyabroad/pinoyachievers/pinoys-worldwide-can-vote-for-jessica-sanchez-through-facebook}. Although we have no proof of any irregular voting activity, tweets analysis clearly points out to a possible anomaly that may be a concern. 

\section{Post-event analysis}

\begin{figure}
\includegraphics[width=1.0\columnwidth,angle=0]{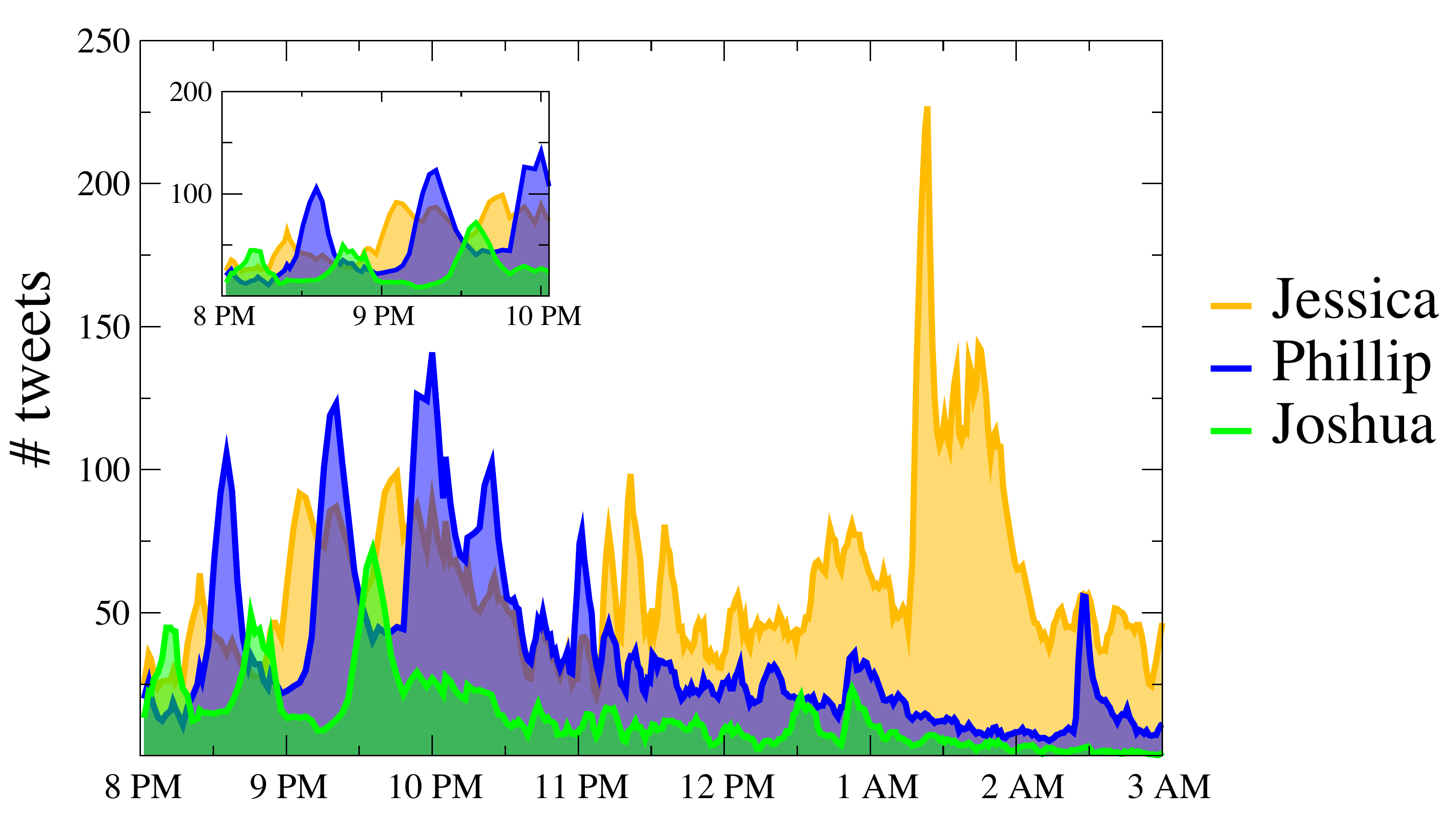}
\caption{Live popularity of the participants to the Top 3 night. The number of tweets related to each one of the Top $3$ contestants is plotted as a function of time, from the start of the show ($8$ PM EST) and the closing of the voting in the West Coast ($3$ AM EST), on Wednesday May $16$, 2012. The data is plotted with the granularity of the minute. The inset magnifies the two hours of the first airing of the show on the East Coast.}
\label{timeseries}
\end{figure}

\begin{table}[t]
  %\begin{ruledtabular} 
    \begin{tabular}{|c|c|c|c|}
    \hline
    { \bf Day} & { \bf Eliminated Cont.} & { \bf Data Indicators} & { \bf Bottom $3$}  \\ \hline %& 
%    $l_e$ & $\langle l \rangle$ & $ \langle \Delta t_l \rangle$ &
%    $\langle l_{proj} \rangle$ \\ \hline %old <s>
      May $17$& Joshua & $\checkmark$ & N/A\\ \hline
      May $10$& Hollie& $\checkmark$ & N/A\\ \hline
      May $3$& Skylar& \checkmark& $\checkmark$ \\ \hline
      April $26$& Elise & CC & $\checkmark$ \\ \hline
      April $19$& Colton & $\times$ & $\checkmark$ \\ \hline
      April $12$& Jessica (saved)& $\times$ &  $(2/3)$\\ \hline
      April $05$&   DeAndre & CC & $\checkmark$   \\ \hline
      March $29$ &  Heejun & CC & $(2/3)$  \\ \hline
      March $22$ & Erika & CC & $(1/3)$  \\ \hline
    \end{tabular}
  %\end{ruledtabular} 
  \caption{We consider the last nine eliminations. In the table we report the date of the elimination, the contestant eliminated, whether the data indicators correctly single out the elimination ($\checkmark$), it is wrong ($\times$) or provide a to close to call (CC). In the last column we compare the data indications for the bottom three (two) contestants announced during the first seven eliminations. We report when within error bars the signal identifies  the bottom three  contestants ($\checkmark$), two out of three  ($2/3$) or one out of three ($1/3$) contestants. } \label{tab:summary}
\end{table}

\begin{figure}[!t]
\includegraphics[width=1.0\columnwidth,angle=0]{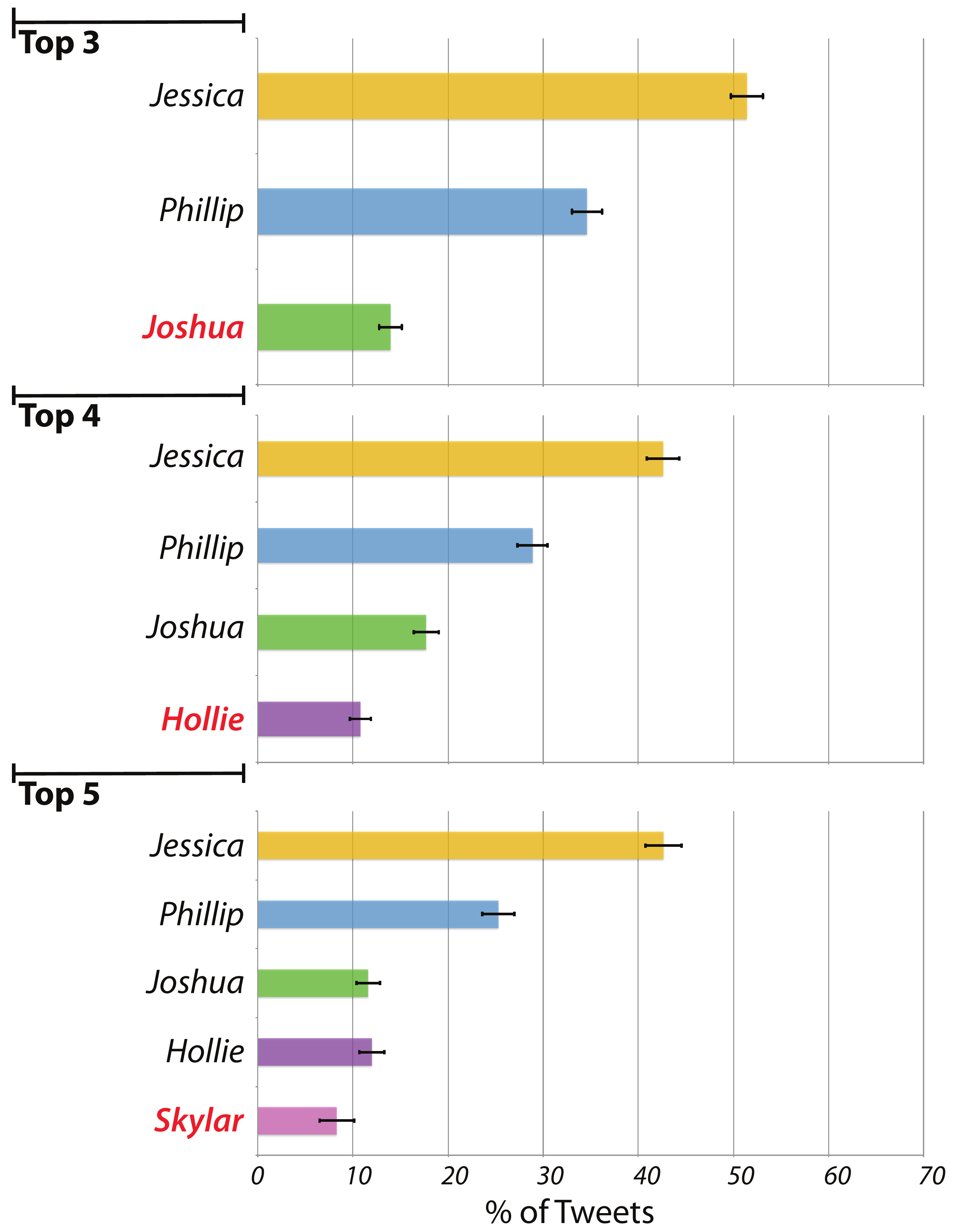}
\caption{For the last three eliminations we plot the ranking of each contestant measured as the percentage of tweets in the time window $8.00$PM -$3.00$AM EST of the last three Wednesdays. We plot the $99\%$ confidence intervals. In red we mark the contestant that was eliminated the next day. It is clear that even considering the errors, the ranking done considering the volume of tweets related to a specific contestant, is sufficient to identify the least preferred.  }
\label{rank}
\end{figure}

Our fundamental, and somehow naive, assumption is that the number of votes each contestant receives is proportional to the number of tweets that mention her. In other words, the larger the number of tweets  referred to a contestant - the twitter volume - the larger the number of votes she will get.  This gives a natural measure to rank each contestant. It is important to note that this is a very simple measure, and that we deliberately choose not to take into account many of the factors that in principle might affect the results, such as the presence of negative or neutral tweets, or attempts to directly affect the counts by spamming the system with automatically generated tweets. In fact, one of the goals of this paper is to test whether or not a minimal set of  measures applied to Twitter data can be good indicators of the actual voting outcome. Past attempts have met with ambivalent results and we are interested in testing the limits of this naive approach by building an unsophisticated prediction system assembled in less than one week.

While our dataset spans the entire duration of the current season, we focus only on the top-ten phase of the show, when just $10$ contestants remained and test the predictive power of the Twitter proxy against the last $9$ eliminations. For $7$ of those, the "bottom-three" contestants, the least three voted contestants ($2$ in the elimination of May 3rd )  were revealed during the iconic part of the show: elimination day. We consider not just the success in predicting the contestant that will be eliminated but also the three that received the least votes. \\
In order to minimize the noise that might be introduced by discussions after the voting time and especially after the elimination, we considered the number of tweets generated on a specific time window: $8.00$ PM - $3.00$ AM EST each Wednesday. The show airs at $8$ PM EST. The votes can be submitted until midnight in the West coast which translates to $3.00$ AM in the east. In Figure~\ref{timeseries} we show the number of tweets related to each of the top three contestants for every minute of the voting window on May $16$. Interestingly the number of tweets associated to the eliminated contestant (Joshua) is practically always the smallest. The inset provided a detailed view of the live show time period.  At this resolution the sequence of peaks of each contestant correlates with time and sequence of their performances that night. 

For each of the last $9$ weeks, we have integrated the number of tweets related to each user in the show+voting time window. We then ranked the  contestants in decreasing order. The last $3$ count as the bottom three and the last contestant is the most likely to be eliminated. We confront our prediction with the real outcomes. To account for errors induced by sampling of the real number of tweets we evaluated the $99\%$ confidence intervals assuming a homogenous and fair sampling and report the results in Table~\ref{tab:summary} . Twitter data serves as a correct indicator for the last three eliminations and identifies correctly most of the bottom three/two contestants. 

Twitter signal indications were wrong two times, and we have other four cases in which the confidence intervals in the ranking could not allow to make a prediction (too close too call). In Figure~\ref{rank} it is possible to notice that, as expected, when the number of contestants reduces and the fan base solidifies, the differences between ranks become much clearer and separated.  

\section{And the winner is...}
\begin{figure}[!t]
\includegraphics[width=1.0\columnwidth,angle=0]{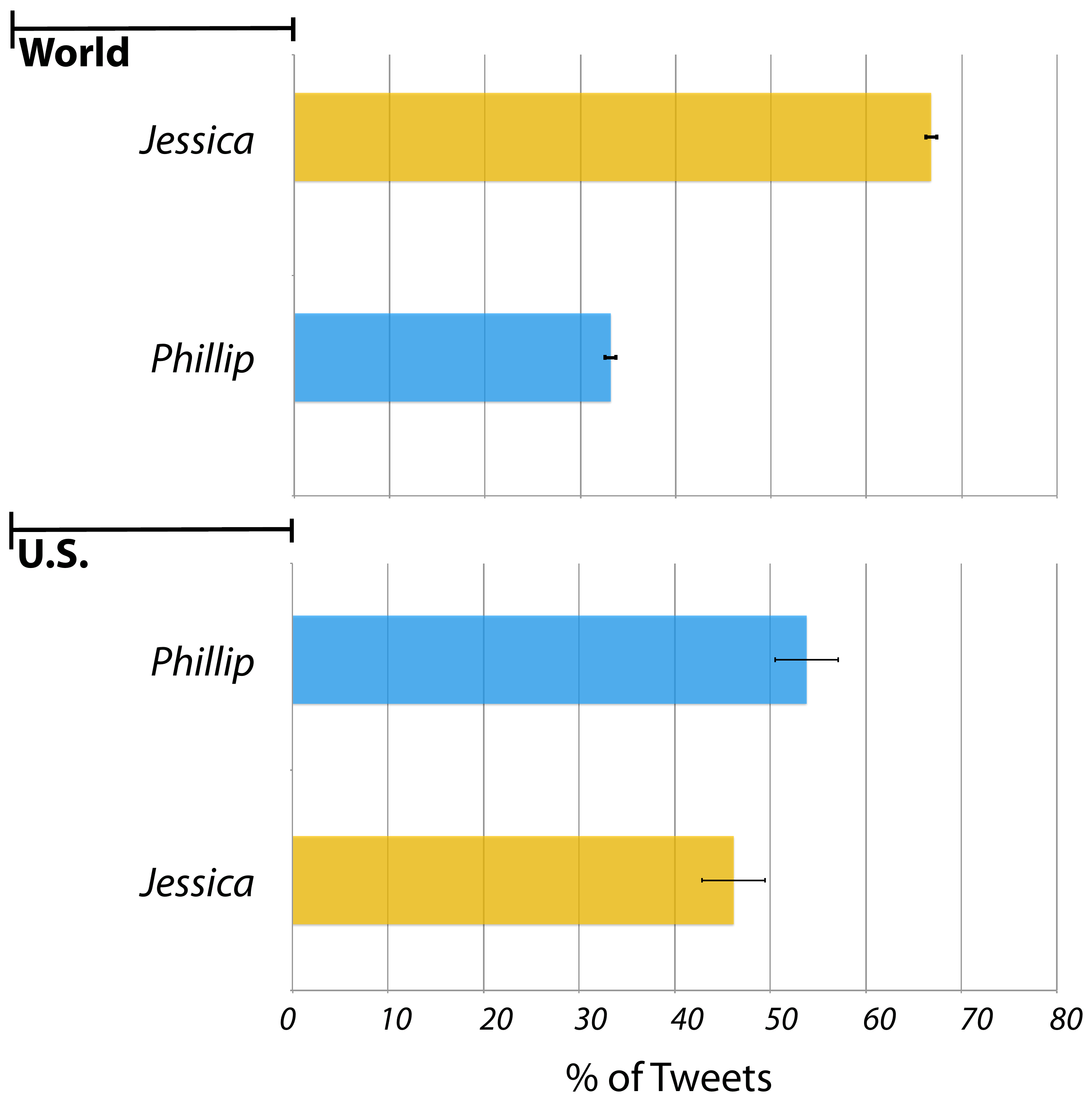}
\caption{Finale ranking. The ranking of the two contestants of season finale, measured as the percentage of tweets in the time window $8.00$PM-$4.00$AM EST, is plotted. The top histogram takes in to account the whole dataset (World), while the bottom one only considers the set of tweets geolocalized in the United States (U.S.). We report the $99\%$ confidence intervals.  }
\label{rank_finale}
\end{figure}

The analysis of the season finale is based on the data collected between the beginning of the show in the East at $8.00$ P.M. EST and the end of the voting period in the West, at $4.00$ A.M. EST. The histogram of Figure~\ref{rank_finale} has a twofold interpretation. If we consider the whole of our dataset, as we have done in the previous analysis, Jessica turns out to have been the most popular in Twitter in our time window. Henceforth, the analysis analysis used for the elimination shows  lead us to predict that Jessica will be the winner of the show. 

\begin{figure}
\includegraphics[width=0.9\columnwidth,angle=0]{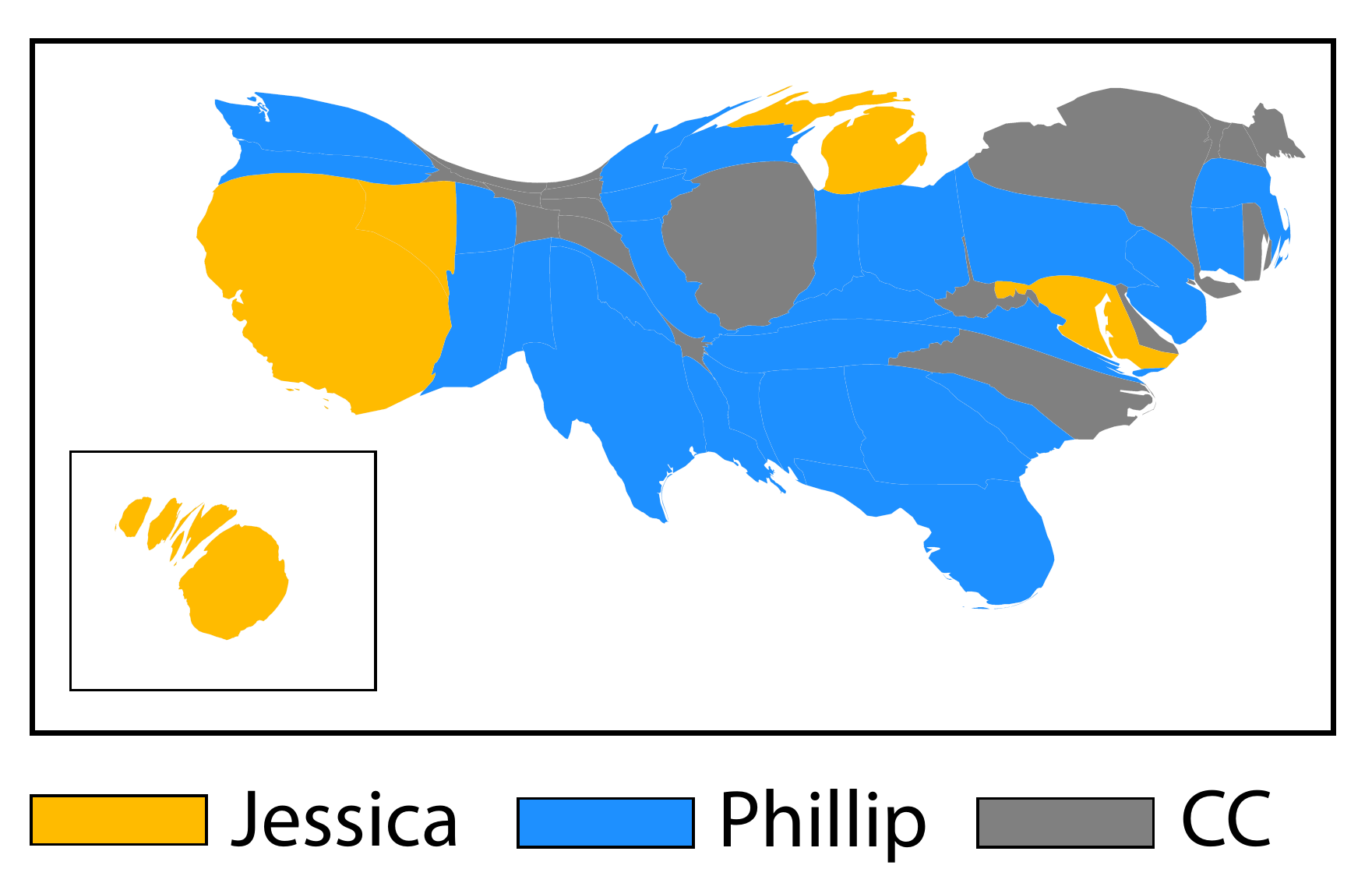}
\caption{Geographical polarization of the tweets for the Top $2$ contestants. The area of each State is proportional to the number of geolocalized tweets generated there, while the color represents the contestant with the majority of the vote. The grey represent states we could not assign to a single contestant within the statistical errors (CC). }
\label{carto_finale}
\end{figure}

However, there is an important caveat. As we pointed out before, Jessica is the only contestant that has a strong Twitter signal originating from outside of the U.S. (and in particular from the Philippines), with an increasing trend after the show on April $19$. Given that the voting is restricted to the U.S. only, it is helpful to have a closer look at the data, and consider the subset of Tweets that come with geographical metadata. Although the geolocalized data are a much smaller subset of the total signal, this dataset allows us to provide the contestants' standing  restricted to the USA Twitter population. In the US, Phillip appears to have the largest fanbase of the two contestants (see also the cartogram of Figure~\ref{carto_finale}). If the possibility of votes coming from abroad is discarded, using the available data, we could then claim that Phillip is going to be the winner of the $11th$ edition of American Idol. However, the data show that the advantage of Phillip in the U.S. is remarkably smaller than the one of Jessica in the aggregated dataset, and the voting coming from abroad might have a crucial role in determining the outcome of the finale.

\section{Conclusion}

We have shown that the open source data available on the web can be used to make educated guesses on the outcome of societal events. Specifically, we have shown that extremely simple measures quantifying the popularity of the American Idol participants on Twitter strongly correlate with their performances in terms of votes. A post-event analysis shows that the less voted competitors can be identified with reasonable accuracy (Table~\ref{tab:summary}) looking at the Twitter data collected during the airing of the show and in the immediately following hours.

It is worth noting that our analysis aims to be extremely simple in order to establish a valid baseline on what it is possible to deduce by Social Media. As such, we purposefully do not consider a number of refinements and techniques that could improve the accuracy of our predictions. Distortions due to overactive users can be controlled by evaluating the number of unique users tweeting on each contestant. The text of the tweets could be scrutinized by using sentiment analysis techniques to select and compare only specific positive or negative tweets as a proxy for success/failure. Corrections to the demographic representations of Twitter users could be considered. All these techniques have been or are being developed in the analysis of a wealth of social phenomena and could be tested in a very clear and simple setting such as those of American Idol or similar shows. 
 
Furthermore, we have illustrated that open source data can provide a deeper insight into the composition of the audience, with the eventual possibility of pointing out possible sources of anomalous behaviors. A geographical projection of the data reveals a non-uniform distribution of the basins of fans, and likely of voters, for the different participants. Interestingly, the same inspection highlights also that a strong activity concerning some of the candidates may come from non-U.S. countries, whose audience are officially forbidden to vote. 
 
Finally, our work casts a word of warning on the possible feedback between competitive TV shows and social media. Indeed, while the former rely more and more on the online voting of the audience, and the votes are kept secret and revealed only at the end of the show, all of the data necessary to monitor and even forecast the outcome of these shows is publicly available on the web. Given the large economic interests that lay behind such programs, such as the revenues of betting agencies and the major contracts of the show participants, it is obvious that this situation can lead to a number of undesirable outcomes. For example, the audience could be induced to alter their behavior in function of the situation they observe, and the job of betting agencies could be dramatically simplified. On a more general basis, our results highlight that the aggregate preferences and behaviors of large numbers of people can nowadays be observed in real time, or even forecasted, through open source data freely available in the web. The task of keeping them private, even for a short time, has therefore become extremely hard (if not impossible), and this trend is likely to become more and more evident in the future years.

\section*{Acknowledgements}

The authors would like to thank Duygu Balcan for generating the cartograms used in this manuscript.

\bibliography{refs}
\end{document}